\documentclass{article}
\usepackage{ijcai26}

\usepackage{times}
\usepackage{soul}
\usepackage{url}
\usepackage[hidelinks]{hyperref}
\usepackage[utf8]{inputenc}
\usepackage[small]{caption}
\usepackage{graphicx}
\usepackage{amsmath}
\usepackage{amsthm}
\usepackage{booktabs}
\usepackage{algorithm}
\usepackage{algorithmic}
\usepackage[switch]{lineno}

\title{Hierarchical Reranking for Scalable Financial RAG System}

\author{
Joohyun Lee$^1$
\and
Sungwoo Hong$^2$
\affiliations
$^1$Financial Security Institute\\
$^2$Hanyang University\\
\emails
dlee110600@gmail.com,
toggiya0701@gmail.com
}

\begin{document}

\maketitle

\begin{abstract}
Analyzing financial documents such as 10-K filings, tabular disclosures, and macroeconomic reports demands expert reasoning and extensive time. However, existing Retrieval-Augmented Generation systems often struggle to process hybrid text–table structures or massive scale of financial documents. To address these challenges, we propose Hierarchical Reranker, a RAG framework designed to improve retrieval performance and generative reliability across large-scale financial datasets. The system integrates three key innovations: Pre-Retrieval Optimization, enhancing query clarity and search efficiency through normalization, keyword expansion, and table transformation; Hierarchical Reranker Architecture, improving retrieval precision through a two-stage ranking mechanism; and Long-Context Management, preserving reasoning accuracy through adaptive input partitioning and fusion under extensive contexts. Across multiple benchmarks, including FinQA, FinanceBench, and ConvFinQA, the proposed system achieved an NDCG@20 score of 0.7918 and demonstrated superior factual consistency. Its robustness was further validated by achieving second place in the ACM-ICAIF '24 FinanceRAG Challenge. This work presents a deployable, domain-optimized RAG pipeline that enhances both the accuracy and scalability of financial reasoning, paving the way for automated audit reporting and quantitative investment analysis. The source code will be made publicly available on GitHub upon acceptance.
\end{abstract}

\section{Introduction}
Interpreting 10-K filings, reconciling tabular disclosures, and contextualizing results against shifting macroeconomic conditions remain among the most labor-intensive activities in the financial industry. These tasks resist automation because they demand both expert domain judgment and exhaustive cross-referencing across long, heterogeneous documents, resulting in high operational cost and limited scalability \cite{Lee2025}. As financial institutions begin to deploy autonomous LLM-based agents for auditing, quantitative research, and portfolio analysis \cite{wu2023bloomberggptlargelanguagemodel,Papasotiriou_2024}, the bottleneck has shifted from \emph{whether} LLMs can read financial text to \emph{how reliably and economically} they can ground their reasoning in the underlying evidence.

Retrieval-Augmented Generation (RAG) is the natural substrate for this grounding \cite{yang2023investlmlargelanguagemodel}, yet off-the-shelf RAG pipelines exhibit three persistent failure modes when transplanted into finance \cite{yu2025tableragretrievalaugmentedgeneration,databricks2024rag}: (1) domain-specific jargon, units, and abbreviations cause query--corpus embedding drift, degrading recall; (2) hybrid text--table evidence is poorly aligned by retrievers trained on prose, leading to numerically incorrect generations; and (3) reasoning quality collapses on long inputs, even on models that nominally support 100k+ token contexts. Each failure mode interacts with the others: a financial 10-K can simultaneously be long, dense in jargon, and dominated by tables.

We address these failures with the Hierarchical Reranker, a finance-specific RAG framework engineered for deployable, large-scale use. Rather than scaling a single monolithic retriever or generator, the framework decomposes the problem across cooperating components, each addressing one failure mode. The two-stage retrieval design in particular instantiates a \emph{small-and-large model collaboration}: a lightweight reranker rapidly prunes the candidate pool, and a high-capacity reranker performs fine-grained semantic adjudication on the survivors. This decomposition keeps end-to-end latency and token cost bounded while concentrating expensive computation where it matters --- a property we view as essential for any RAG system that is to operate at institutional scale.

Our contributions are threefold:
\begin{itemize}
\item \textbf{Pre-Retrieval Optimization.} We normalize finance-specific units and abbreviations, augment queries with domain keywords, and deterministically convert Markdown tables to JSON so that numeric values stay bound to their headers --- producing measurable gains in retrieval quality without introducing LLM-induced hallucinations during preprocessing.
\item \textbf{Hierarchical Reranker.} A two-stage cascade pairs a fast first-stage filter with a high-capacity second-stage reranker, capturing cross-sentence financial dependencies and table--text consistency while keeping inference cost bounded.
\item \textbf{Long-Context Management.} For inputs beyond a 64k-token threshold, we partition evidence into semantically coherent chunks and fuse intermediate answers with an evidence-aware merger that explicitly handles ambiguous or contradictory partials. Ablations isolate the contribution of each component.
\end{itemize}

The framework was validated through extensive ablation studies and benchmark evaluations, has been deployed in institutional auditing and investment workflows, and secured \textbf{2nd place} in the ACM-ICAIF '24 FinanceRAG Challenge~\cite{icaif-24-finance-rag-challenge}, demonstrating both robustness and industry-level competitiveness.

\begin{figure*}[t]
\centering
\includegraphics[width=\textwidth]{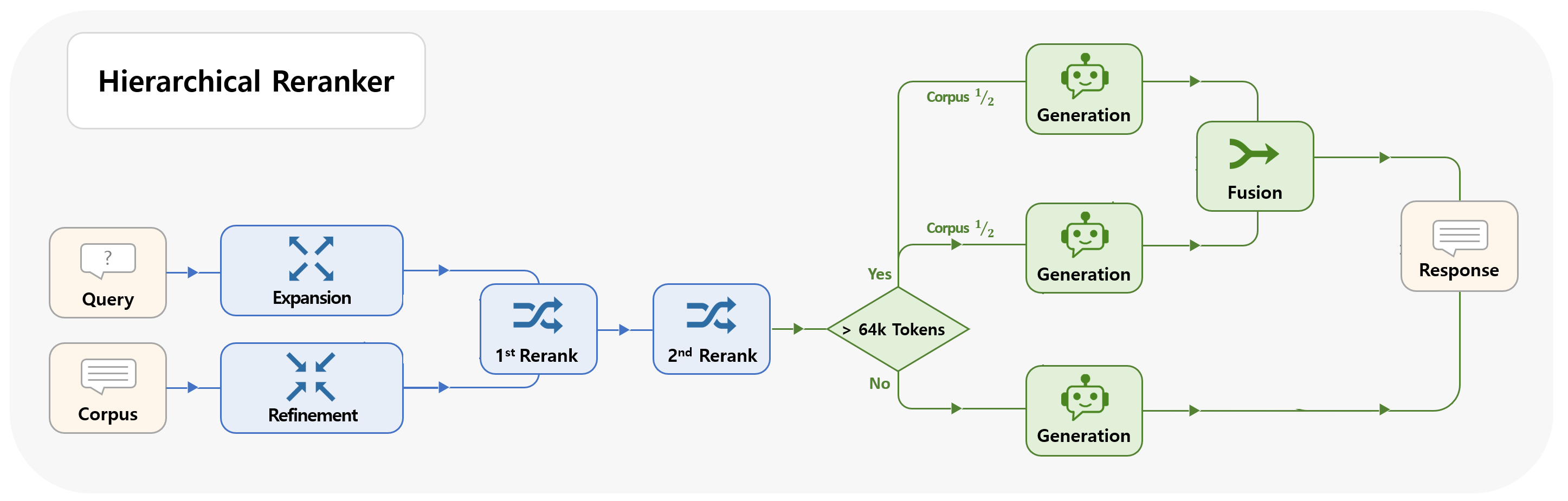}
\caption{Hierarchical Reranker Framework, which integrates query expansion, corpus compression, and a two-stage reranking pipeline to enhance retrieval performance. The retrieved corpora are dynamically managed under a long-context fusion mechanism, ensuring efficient and accurate response generation even for inputs exceeding 64k tokens.}
\label{fig:banner}
\end{figure*}

\section{Related Work}
The integration of artificial intelligence in the financial domain has rapidly advanced through the emergence of retrieval-based benchmarks and reasoning datasets \cite{Lee2025}. Early datasets such as FinQABench \cite{finqabench} and FinanceBench \cite{islam2023financebenchnewbenchmarkfinancial} focused primarily on factual grounding in financial documents like 10-K filings, evaluating models' ability to reduce hallucinations and improve factual correctness. While these benchmarks improved retrieval--generation alignment, they often assumed relatively short contexts and ignored multi-step numerical reasoning.

Subsequently, TATQA \cite{zhu2021tatqaquestionansweringbenchmark}, FinQA \cite{chen2022finqadatasetnumericalreasoning}, and ConvFinQA \cite{chen2022convfinqaexploringchainnumerical} expanded the task scope to hybrid textual--tabular reasoning, requiring models to perform arithmetic and comparative analysis. These datasets highlighted the limitations of general-purpose retrievers and LLMs in understanding structured quantitative data, motivating research into more domain-aware retrieval architectures.

From the retrieval perspective, query expansion \cite{jagerman2023queryexpansionpromptinglarge} and reranking \cite{ma2023zeroshotlistwisedocumentreranking} have been widely studied to enhance search accuracy. Methods such as HyDE \cite{gao2022precisezeroshotdenseretrieval} and query rewriting \cite{liu2024queryrewritinglargelanguage} improved retrieval recall through lexical and semantic enrichment, while rerankers refined document relevance post-retrieval. However, most existing studies \cite{fan2024surveyragmeetingllms} have been optimized for horizontal tasks, focusing on either pre-retrieval or post-retrieval, with few demonstrating performance improvements in domain-specific (vertical) applications.

Long-context reasoning represents another key research stream. Despite advances in models like Claude-Opus 4.6 \cite{claude46}, GPT-5.4 \cite{gpt54}, Grok-4.20 \cite{grok420} and Gemini-3.0-Pro \cite{gemini30}, several studies \cite{databricks2024rag,an2024doeseffectivecontextlength} have observed performance degradation beyond 64k tokens, particularly in financial reasoning tasks with dense numerical references. Existing work \cite{li2023compressingcontextenhanceinference} offers few solutions for dynamically constraining and fusing contexts while preserving reasoning consistency. By integrating pre-retrieval optimization, hierarchical reranker refinement, and context-length management into a unified finance RAG pipeline, our study aims to build a practically deployable RAG system for real-world financial applications.

\section{Tasks and Dataset}
To construct a finance-specific RAG system, we define two interdependent tasks that jointly determine the overall pipeline design: Document Retrieval and Answer Generation.

\subsection{Task 1: Document Retrieval}
Given a user query, the objective is to identify the top 20 most relevant passages from a large corpus of financial documents. Unlike general-purpose retrieval tasks, financial corpora contain both textual and numerical data, making semantic and quantitative alignment equally important. We thus formulate retrieval as a two-stage ranking problem embedding-based coarse retrieval followed by fine-grained reranking to balance efficiency and accuracy. Retrieval performance is evaluated using Normalized Discounted Cumulative Gain (NDCG@20), which captures both ranking quality and semantic relevance. This metric is chosen for its robustness in evaluating graded relevance rather than binary correctness, which aligns well with real-world financial information retrieval.

\subsection{Task 2: Answer Generation}
Once the top-ranked corpora are retrieved, the goal is to generate a factual, numerically grounded answer that directly references the evidence within the selected documents. This task extends beyond text summarization by requiring precise interpretation of tables, ratios, and cross-document dependencies. To assess generation quality, we adopt the LLM-as-a-Judge framework \cite{gu2025surveyllmasajudge}, which compares model outputs with ground-truth answers while evaluating factual consistency, logical reasoning, and numerical precision. This was feasible because the benchmark answers are short and clear (mainly numbers or simple words), allowing evaluation through straightforward accuracy metrics. Although the evaluation relied solely on LLM-as-a-Judge, the deterministic and fact-based nature of the benchmark tasks minimizes heuristic bias, making statistical significance testing less critical.

\subsection{Datasets}
The proposed system is benchmarked on multiple finance-oriented datasets, each emphasizing distinct aspects of retrieval and reasoning. Each dataset sample consists of a natural-language query, a set of chunked corpora, the corresponding ground truth, and document sources. Collectively, these datasets provide a comprehensive benchmark suite for assessing both retrieval effectiveness and numerically grounded generation in real-world financial scenarios.
\begin{itemize}
\item \textbf{FinQABench}: Based on 10-K filings; focuses on detecting hallucinations in generated answers and ensuring factual correctness.
\item \textbf{FinQA}: Derived from earnings reports; evaluates multi-step numerical reasoning using both tabular and textual data.
\item \textbf{ConvFinQA}: Also based on earnings reports; assesses model performance on conversational financial queries.
\item \textbf{FinanceBench}: Built on 10-K filings; measures a system's ability to handle real-world financial questions with domain precision.
\item \textbf{TATQA}: Composed of financial reports; tests arithmetic, comparative, and logical reasoning over hybrid tabular--text data.
\end{itemize}

\section{Method}
The proposed financial RAG system is composed of three sequential stages. The first two stages correspond to Task 1 (Retrieval), while the third stage corresponds to Task 2 (Generation). Overall, the pipeline is divided into three core phases: Pre-Retrieval, Retrieval, and Generation. As illustrated in Figure~\ref{fig:banner}, the blue components represent the retrieval process, while the green components correspond to generation.

\subsection{Pre-Retrieval}
The Pre-Retrieval phase is designed to enhance the interpretability of financial queries and the consistency of the document corpus before embedding-based retrieval. Financial texts often contain abbreviations, implicit relations, and domain-specific terminology, which often lead to semantic mismatches between queries and document embeddings. To mitigate these issues, we propose a four-stage pre-processing pipeline that improves both query clarity and corpus normalization.

\subsubsection{Normalization}
All input queries are first normalized to reduce lexical ambiguity. This step includes lowercase, typographical correction, and expansion of financial abbreviations (e.g., EPS: Earnings per share, YoY: Year-over-Year). Measurement units such as ``K'' or ``M'' are standardized into consistent numeric expressions, and missing contextual terms are restored based on document metadata. This ensures that semantically equivalent expressions share a uniform representation within the embedding space. The corpus is normalized likewise for consistency.

\subsubsection{Keyword Extraction}
Financial-specific keywords are extracted and appended to each query to strengthen the alignment between query and corpus embeddings. This increases the density of financial terminology within the input, allowing the retriever to better capture hybrid text--table semantics commonly found in financial documents.

\subsubsection{Paraphrasing}
Each normalized query is semantically expanded through paraphrasing. This step generates multiple linguistically diverse but semantically equivalent variants, enabling the retriever to generalize across different formulations of financial questions (e.g., ``What is Apple's revenue in 2023?'' vs. ``Report Apple's 2023 revenue''). This expansion broadens the search space without introducing significant computational overhead.

\subsubsection{Hypothetical Document Generation (HyDE)}
Inspired by HyDE-based methods, a synthetic pseudo-document is generated for each query to represent a plausible context where the answer could appear. These hypothetical passages act as semantic anchors that enhance embedding alignment between the abstract query and the domain-specific corpus.

\subsubsection{Table-to-json}
During the corpus pre-processing stage, it is crucial to maintain semantic coherence across documents while ensuring the accuracy of numerical information. Accordingly, each document was semantically chunked at the sentence level to prevent contextual fragmentation. In addition, since large-scale tables in financial documents contain key quantitative information, Markdown-style tables were converted into JSON structures (table-to-json) using a rule-based script, rather than relying on LLM-based conversion, to prevent potential hallucinations.

This process explicitly preserves the relationships between numerical values and their headers, thereby strengthening the semantic alignment between textual and numerical data during the embedding and retrieval stages. The effectiveness of this transformation was validated through comparative experiments between the Markdown format (original tables) and the JSON format (table-to-json).

\subsubsection{Summary}
When the corpus size is extremely large (over 10k tokens), utilizing all documents as-is may be inefficient and could lower retrieval performance. To address this, each document was replaced through summarization substitution, preserving only the essential financial indicators, results, and contextual information. This compression removes unnecessary narrative sentences while prioritizing semantically central statements, thereby improving both the efficiency and accuracy of the retrieval stage.

Most Pre-Retrieval steps were performed using Claude-Opus-4.6 \cite{claude46}, while certain quantitative tasks, such as table conversion, were executed through rule-based scripts. Through this bidirectional normalization and structuring process applied to both queries and corpora, the system maximizes semantic compatibility between complex textual--numerical data in the financial domain and enables more precise and reliable retrieval in subsequent stages.

\subsection{Retrieval}
The Retrieval stage identifies the most semantically and numerically relevant passages from a large-scale financial corpus, bridging the preprocessed query and the downstream generator. The design objective is twofold: maximize retrieval precision on hybrid text--table evidence, and bound inference cost so that the system remains deployable at institutional scale.

We therefore avoid a single monolithic reranker and instead employ a two-stage hierarchical reranker in which a small, fast model and a large, accurate model cooperate. Each model is specialized for the regime where it dominates: the small model handles coarse lexical pruning across the full candidate pool, while the large model performs deep semantic adjudication on a much smaller surviving set. This division of labor decouples \emph{breadth} from \emph{depth} and concentrates expensive computation where it has the highest marginal value.

\subsubsection{Stage 1: Lightweight Filtering}
In the first stage, a fast, low-complexity jina-reranker-v3 \cite{jinarerankerv3} is used to eliminate low-relevance or noisy candidates. This model is optimized for lexical and shallow semantic similarity, leveraging extended context support up to 131k tokens. By narrowing the candidate pool to the top 100 passages, it significantly reduces downstream computational load without sacrificing recall.

\subsubsection{Stage 2: Fine-Grained Semantic Reranking}
The second stage applies a high-capacity reranker to re-evaluate the top 100 candidates and extract the final top 20 passages. This reranker focuses on fine-grained contextual relationships, such as cross-sentence financial dependencies and numerical consistency across tables and text. Through this hierarchical refinement, the system effectively captures both semantic and quantitative correspondence, which is crucial for hybrid financial corpora.

This architecture achieves an optimal trade-off between precision and efficiency. The lightweight first stage prevents unnecessary computation on irrelevant candidates, while the second stage provides the semantic depth needed to identify financially meaningful evidence. This separation of lexical filtering and contextual reasoning enables robust retrieval performance even under high-volume workloads, making the approach scalable for institutional use.

Financial documents often combine narrative text and tabular disclosures, requiring models to align textual descriptions with structured numerical information. To handle this, the retrieval module leverages both the normalized corpus (from the Pre-Retrieval phase) and the JSON-formatted tabular data, allowing the reranker to compute cross-modal similarity between natural language and numeric fields. This design improves factual consistency and ensures that retrieved contexts are suitable for quantitative reasoning.

The final output consists of the top 20 ranked passages, which collectively form a compact yet information-rich context for the generation stage. This cap balances semantic coverage and token efficiency, ensuring that subsequent long-context generation operates within model input limits while retaining all essential evidence.

\begin{algorithm}[t]
\caption{Proposed Framework}
\label{financerag_alg}
\textbf{Input}: Query $Q$, Corpus $C$\\
\textbf{Output}: Response $R$
\begin{algorithmic}[1]
\STATE $Q' \gets \text{Normalization}(Q) \cup \text{Keywords-Extraction}(Q)$
\STATE $C' \gets \text{Normalization}(C) \cup \text{Table2Json}(C)$
\STATE $C_{100} \gets \text{Rerank}_1(Q',\, C')$
\STATE $C_{20} \gets \text{Rerank}_2(Q',\, C_{100})$
\IF{$\text{tokens}(Q' \cup C_{20}) \le 64\mathrm{k}$}
  \STATE $R \gets \text{LLM}(Q',\, C_{20})$
\ELSE
  \STATE $R_1 \gets \text{LLM}(Q',\, C_{1{:}10})$
  \STATE $R_2 \gets \text{LLM}(Q',\, C_{11{:}20})$
  \STATE $R \gets \text{Fusion}(R_1,\, R_2)$
\ENDIF
\STATE \textbf{return} $R$
\end{algorithmic}
\end{algorithm}

\begin{table*}[t]
\centering
\caption{Ablation study of Pre-Retrieval components: Norm denotes normalization of queries and corpora, including abbreviation expansion, unit standardization, and typo or grammar correction. HyDE represents Hypothetical Document Embedding, where a pseudo-document is generated to enhance semantic alignment between the query and the corpus.}
\label{tab:pre-retrieval}
\begin{tabular}{lcccc|cccc|c}
\toprule
\multicolumn{5}{c|}{\textbf{Query}} & \multicolumn{4}{c|}{\textbf{Corpus}} & \textbf{NDCG@20} \\
\cmidrule(lr){1-5} \cmidrule(lr){6-9}
Original & Norm & Keywords & Paraphrased & HyDE & Original & Norm & Table-to-json & Summary &  \\
 &  & Extraction &  &  &  &  &  &  &  \\
\midrule
$\circ$ & - & - & - & - & $\circ$ & - & - & - & 0.7323 \\
- & $\circ$ & - & - & - & $\circ$ & - & - & - & 0.7446 \\
- & $\circ$ & $\circ$ & - & - & $\circ$ & - & - & - & 0.7542 \\
- & $\circ$ & - & $\circ$ & - & $\circ$ & - & - & - & 0.7347 \\
- & $\circ$ & - & - & $\circ$ & $\circ$ & - & - & - & 0.7347 \\
\midrule
$\circ$ & - & - & - & - & - & $\circ$ & - & - & 0.7211 \\
- & $\circ$ & - & - & - & - & $\circ$ & - & - & 0.7333 \\
- & $\circ$ & $\circ$ & - & - & - & $\circ$ & - & - & 0.7503 \\
- & $\circ$ & - & $\circ$ & - & - & $\circ$ & - & - & 0.7446 \\
- & $\circ$ & - & - & $\circ$ & - & $\circ$ & - & - & 0.6759 \\
\midrule
$\circ$ & - & - & - & - & - & $\circ$ & $\circ$ & - & 0.7301 \\
- & $\circ$ & - & - & - & - & $\circ$ & $\circ$ & - & 0.7529 \\
\textbf{-} & \textbf{$\circ$} & \textbf{$\circ$} & \textbf{-} & \textbf{-} & \textbf{-} & \textbf{$\circ$} & \textbf{$\circ$} & \textbf{-} & \textbf{0.7918} \\
- & $\circ$ & - & $\circ$ & - & - & $\circ$ & $\circ$ & - & 0.7677 \\
- & $\circ$ & - & - & $\circ$ & - & $\circ$ & $\circ$ & - & 0.6843 \\
\midrule
$\circ$ & - & - & - & - & - & - & - & $\circ$ & 0.6544 \\
- & $\circ$ & - & - & - & - & - & - & $\circ$ & 0.6501 \\
- & $\circ$ & $\circ$ & - & - & - & - & - & $\circ$ & 0.6579 \\
- & $\circ$ & - & $\circ$ & - & - & - & - & $\circ$ & 0.6542 \\
- & $\circ$ & - & - & $\circ$ & - & - & - & $\circ$ & 0.5853 \\
\bottomrule
\end{tabular}
\end{table*}

\begin{table}[t]
\centering
\caption{Comparison of Hierarchical Reranker Combinations}
\label{tab:reranker-comparison}
\begin{tabular}{lll}
\toprule
\textbf{1\textsuperscript{st}} & \textbf{2\textsuperscript{nd}} & \textbf{NDCG@20} \\
\textbf{Reranker} & \textbf{Reranker} & \\
\midrule
                  & -                      & 0.7260 \\
                  & Linq-Embed-Mistral     & 0.7378 \\
jina-reranker-v3  & gte-Qwen2-7B-instruct  & 0.7567 \\
                  & Qwen3-Reranker-4B      & 0.7763 \\
                  & \textbf{Qwen3-Reranker-8B} & \textbf{0.7918} \\
\bottomrule
\end{tabular}
\end{table}

\subsection{Generation}
Although recent LLMs nominally accept extremely long inputs \cite{gpt54,claude46}, multiple studies \cite{databricks2024rag,jin2024long,paulsen2025contextneedmaximumeffective} report that response quality, and in particular numerical fidelity, degrades well before the advertised context limit. The gap between \emph{nominal} and \emph{effective} context length is especially consequential in finance, where 10-K filings routinely exceed 100k tokens and where a single misread cell can invalidate downstream reasoning. The generation stage must therefore decide not how much context to feed the model, but how to feed it in a way that preserves accuracy.

\subsubsection{Context Size Management}
To empirically identify a reliable operational threshold, we conducted ablation studies using several state-of-the-art LLMs under long-context and numerically intensive conditions. We observed a noticeable decline in performance when input size exceeded 64k tokens. Based on this observation, we set 64k as the context threshold for all subsequent experiments.

\subsubsection{Fusion}
When the combined input exceeds 64k tokens, the top-20 corpora are divided into two semantically coherent subsets, producing interim answers $R_1$ and $R_2$. The two outputs are then merged through a conditional fusion process: the system first identifies whether each interim response contains a definitive answer; if only one does, that answer is directly adopted; if both contain valid answers, the one with the higher confidence value is selected; and if neither provides a clear answer, the model outputs an explicit ``unknown'' response. This adaptive fusion ensures consistent and interpretable generation under long-context conditions while preventing hallucinated synthesis across partitions.

\begin{table}[t]
\centering
\caption{Comparison of LLM Performance with/without Context Management}
\label{tab:llm-cm-acc}
\begin{tabular}{lll}
\toprule
\textbf{LLM} & \textbf{Context} & \textbf{Accuracy} \\
             & \textbf{Management} &  \\
\midrule
Gemini 3.0 Pro & -         & 0.7593 \\
               & $\circ$   & 0.7610 \\
\midrule
GPT-5.4        & -         & 0.7786 \\
               & $\circ$   & 0.7794 \\
\midrule
Grok-4.20      & -         & 0.7901 \\
               & $\circ$   & 0.7938 \\
\midrule
Claude-4.6 Opus & -        & 0.8103 \\
                & $\circ$  & \textbf{0.8152} \\
\bottomrule
\end{tabular}
\end{table}

\section{Results}
This section presents the experimental results of the proposed framework, including analyses of Pre-Retrieval design, Hierarchical Reranker architecture, and Long-Context Management.

\subsection{Pre-Retrieval Design}
Table~\ref{tab:pre-retrieval} summarizes the retrieval performance under various combinations of query and corpus preprocessing methods. The results clearly show that normalization yields the highest performance gain. Combining Query Normalization, Keyword Extraction, and Table-to-json Conversion achieved the best score (NDCG@20 = 0.7918), outperforming the baseline (no Pre-Retrieval) by +5.9\%. This confirms that pre-retrieval normalization significantly improves semantic coherence and retrieval performance within financial documents.

\subsection{Impact of Hierarchical Reranker}
Table~\ref{tab:reranker-comparison} compares different reranking combinations. Using \textit{jina-reranker-v3} as the lightweight first-stage model followed by \textit{Qwen3-Reranker-8B} \cite{qwen3reranker} as the second-stage model achieved the best performance (NDCG@20 = 0.7918). This hierarchical reranking improved relevance by +6.5\% compared to a single-reranker setup. By separating a lightweight model for fast filtering (Stage1) and a larger model for fine-grained reranking (Stage2), the approach aimed to mitigate the inference time limitation while maintaining retrieval performance.

\subsection{Long-Context Management}
Table~\ref{tab:llm-cm-acc} presents the impact of the proposed context segmentation and fusion strategy.
Across all tested LLMs, applying the split-and-fusion mechanism resulted in a slight accuracy improvement ranging from 0.08\% to 0.49\%.
Claude-4.6 Opus achieved the highest performance (Accuracy = 0.8152) under the 64k-token threshold, indicating that context segmentation had only a marginal effect on overall accuracy.

\subsection{Summary}
Overall, the proposed Hierarchical Reranker framework showed consistent and statistically stable improvements in both retrieval and generation performance.
These results can be attributed to three key components:
(1) optimized pre-retrieval algorithms,
(2) a hierarchical reranker architecture that enhances retrieval precision, and
(3) long-context management for stable reasoning across extended inputs.
The system provides a reliable, scalable, and domain-adaptive solution, supporting its applicability to financial document analysis.

\section{Discussion}
This section discusses the strengths, limitations, and future directions of the proposed finance-specific RAG system. We first highlight how each component contributes to the system's effectiveness in real-world financial tasks, then outline key computational and scalability limitations, and finally present potential avenues for improvement and research extension.

\subsection{Contributions}
Three complementary components --- Pre-Retrieval Optimization, the Hierarchical Reranker, and Long-Context Management --- jointly produced consistent gains across retrieval precision, factual consistency, and reasoning stability. The Pre-Retrieval phase is the simplest yet highest-leverage stage: deterministic normalization, keyword augmentation, and Markdown-to-JSON table conversion lifted NDCG@20 by +5.9\% over the no-preprocessing baseline, and did so without introducing LLM-induced hallucinations during preprocessing. This kind of conservative, rule-based engineering is what makes the system reproducible enough for regulated environments such as auditing and investment research.

The hierarchical reranker complements this by realizing a small-and-large model collaboration: the lightweight first-stage model bounds compute, while the high-capacity second-stage model concentrates effort on the candidates most likely to matter. The two-stage cascade improved NDCG@20 by +6.5\% over a single-reranker baseline while keeping per-query inference cost compatible with institutional throughput. We see this pattern --- specialize small models for breadth and large models for depth --- as broadly applicable to other vertical RAG settings.

\subsection{Limitations}
Despite its effectiveness, the system entails several limitations that warrant attention.
First, the hierarchical reranking architecture, while improving precision, introduces additional computational overhead compared to single-stage rerankers.
This cost can be mitigated through user-guided search constraints — for instance, extracting company names from the user query and using them to substantially narrow the scope of candidate documents before reranking is invoked.
Second, the reliance on a fixed 64k-token context threshold, though empirically validated, restricts scalability when processing extremely long financial reports or multi-document reasoning tasks.
Finally, while the framework was evaluated on multiple financial benchmarks, further validation across multilingual or real-time financial streams remains an open challenge for industrial deployment.

\subsection{Future Work}
We see three natural extensions. First, \textbf{dynamic context prioritization} \cite{ikram2025ascendradynamicrequestprioritization} would replace the current fixed 64k threshold with importance-weighted allocation of model attention, reducing information loss during segmentation and fusion on very long filings. Second, we plan to push the small-and-large model collaboration toward \textbf{adaptive, query-conditional reranking}: invoke the heavy second-stage reranker only when the first stage exhibits low confidence, further compressing token spend on routine queries while preserving accuracy on hard ones. Third, embedding the pipeline inside a broader \textbf{agentic workflow} --- with expert-in-the-loop feedback signals and tool use for numerical verification --- could close the remaining gap between retrieval-grounded answers and analyst-grade financial reasoning, moving the system from a single-turn RAG pipeline toward an autonomous financial-analysis agent.

\section{Conclusion}
Financial document analysis sits at the intersection of high economic value and high expertise cost, making it one of the most natural targets for LLM- and RAG-based automation. Global financial institutions \cite{wu2023bloomberggptlargelanguagemodel,Papasotiriou_2024} have already begun integrating retrieval-grounded LLMs into their operational workflows, but the gap between research-grade RAG and institutional-grade deployment remains substantial.

This work narrows that gap. We presented Hierarchical Reranker, a finance-specific RAG framework that couples deterministic Pre-Retrieval Optimization, a small-and-large model Hierarchical Reranker, and adaptive Long-Context Management into a single deployable pipeline. Comprehensive ablations show that each component contributes measurable gains, and the system as a whole achieved NDCG@20 = 0.7918 and secured 2nd place in the ACM-ICAIF '24 FinanceRAG Challenge. Just as importantly, the framework is now running inside real auditing and investment workflows, supporting the claim that careful engineering --- not only scale --- is what carries RAG into production.

We hope this study offers a concrete reference point for vertical RAG design in finance and contributes to the broader trajectory of LLM-based, agentic systems for high-stakes financial reasoning.

\bibliographystyle{named}
\bibliography{finllm-ijcai26}

\end{document}